\documentclass[reprint,amsmath,amssymb,aps]{revtex4-1}
\pdfoutput=1

\usepackage{graphicx}
\usepackage{dcolumn}
\usepackage{bm}

\begin{document}

\preprint{APS/123-QED}

\title{Enhanced sensing of  non-Newtonian effects at ultrashort range with exceptional points in
optomechanical systems}
\thanks{}%

\author{Jian Liu}
 \altaffiliation[]{}
 \author{Lei Chen}
 \altaffiliation[]{}
\author{Ka-Di Zhu}%
 \email{zhukadi@sjtu.edu.cn}
\affiliation{Key Laboratory of Artificial Structures and Quantum Control (Ministry of
Education), School of Physics and Astronomy, Shanghai Jiao Tong
University, 800 DongChuan Road, Shanghai 200240, China,
Collaborative Innovation Center of Advanced Microstructures, Nanjing, China
}%

\date{\today}

\begin{abstract}
We propose an optomechanical nano-gravimeter based on exceptional
points. The system is a coupled cavity optomechanical system, in which the
gain and loss are applied by driving the cavities with a blue detuned and
red detuned electromagnetic field, respectively. When the gain and loss
reach a balance, the system will show the degeneracy of exceptional points,
and any perturbation will cause an eigenfrequencies split, which is
proportional to the square root of the perturbation strength. Compared with
the traditional optomechanical sensors, the sensitivity is greatly enhanced.
This work paves the way for the design of optomechanical ultrasensitive
force sensors that can be applied to detect non-Newtonian effects,
high-order weak interactions, and so on.

\end{abstract}

\pacs{Valid PACS appear here}
\maketitle


\section{INTRODUCTION}

In quantum mechanics and quantum optics, most systems are open and exchange
energy or particles with the surrounding environment. The dynamics of this
open quantum system can be described by non-Hermit Hamiltonian with complex
eigenvalues. It has been proved that parity-time(PT) symmetrical
non-Hermitian Hamiltonian can also have a real eigenvalue spectrum[1],
providing a new design scheme for devices with novel functions[2-4].
Depending on the real or complex nature of the eigenvalues of the
Hamiltonian, the system can change from PT-symmetric to PT-broken phase at
the exceptional points (EPs).The physical existence of EPs has been proven
through many experiments in some different systems, for example,
nanomechanical resonator[5], optical microcavities[6], photonic lattices[7].
In such systems, two or more eigenmodes coalesce at the EPs, leading to a
variety of unconventional effects observed in experiments, such as
loss-induced transparency[8], unidirectional lasing[9], topological
chirality[10], exotic topological states[11], and chaos[12].

Recently, it has been shown that the complex-square-root topology near the
second-order EPs can enhance the sensitivity of an optical system to the
external perturbation[13,14]. If a higher-order EP is used (greater than the
second-order), this can in principle further amplify the frequency shift
induced by the perturbations, resulting in higher sensitivity[15]. A growing
number of theoretical and experimental studies have been devoted to such
applications[16,17].

On the other hand, some recent theories of physics beyond the standard model
point to the possibility of new physics related to gravity. Those theories
that attempt to unify the standard model with gravity, predict the existence
of extra dimensions, exotic particles, or light moduli in string theory that
could cause a mass coupling in addition to the Newtonian gravitational
potential at short distances[18]. The Yukawa potential due to new
interactions is typically taken to modify the gravitational inverse square
law

\begin{equation}
V(r)=-G_{N}\frac{m_{t}m_{s}}{r}(1+\alpha e^{-r/\lambda }),
\end{equation}%
where $G_{N}$ is Newton's constant, $m_{t}$ and $m_{s}$ are the test mass
and source mass respectively, $r$ is the center of mass separation, and $%
\alpha $ is the strength of any new effect with a length scale of $\lambda $%
. Many experimental and theoretical schemes are dedicated to detecting
gravity deviations at short distances[19-22]. Among these different systems,
benefiting from the high mechanical quality factor properties and the
all-optical-domain methods without heat effect and energy loss caused by
circuits, in recent years, the nano-optomechanical systems have played an
important role in the application of quantum limit detection, including
force[23,24], mass[25], displacement[26] and so on.

However no matter which system is used, force-sensing are mainly based on
the mechanism of linear dependent between the force and displacement[23] or
the force and the resonance frequency[24]. Here we demonstrate an
alternative force-sensing scheme that can operate at non-Hermitian spectral
degeneracies (known as the exception point), thereby increasing the
sensitivity of the optical system. Our proposal is based on two mechanically
coupled optomechanical resonators which are placed in two optical cavities
respectively. The cavities are driven with blue-detuned and red-detuned
laser, respectively. At the balance between gain and loss, the system
features an EP in its mechanical spectrum. Any variation of force gradient
will perturb the system from its exceptional point, leading to frequency
splitting. The sensitivity enhancement more than $10^{6}$\ was theoretically
demonstrated with a sensor operating at exceptional points.

\section{THEORY FRAMEWORK}

\begin{figure}[tbp]
\includegraphics[width=8.5cm]{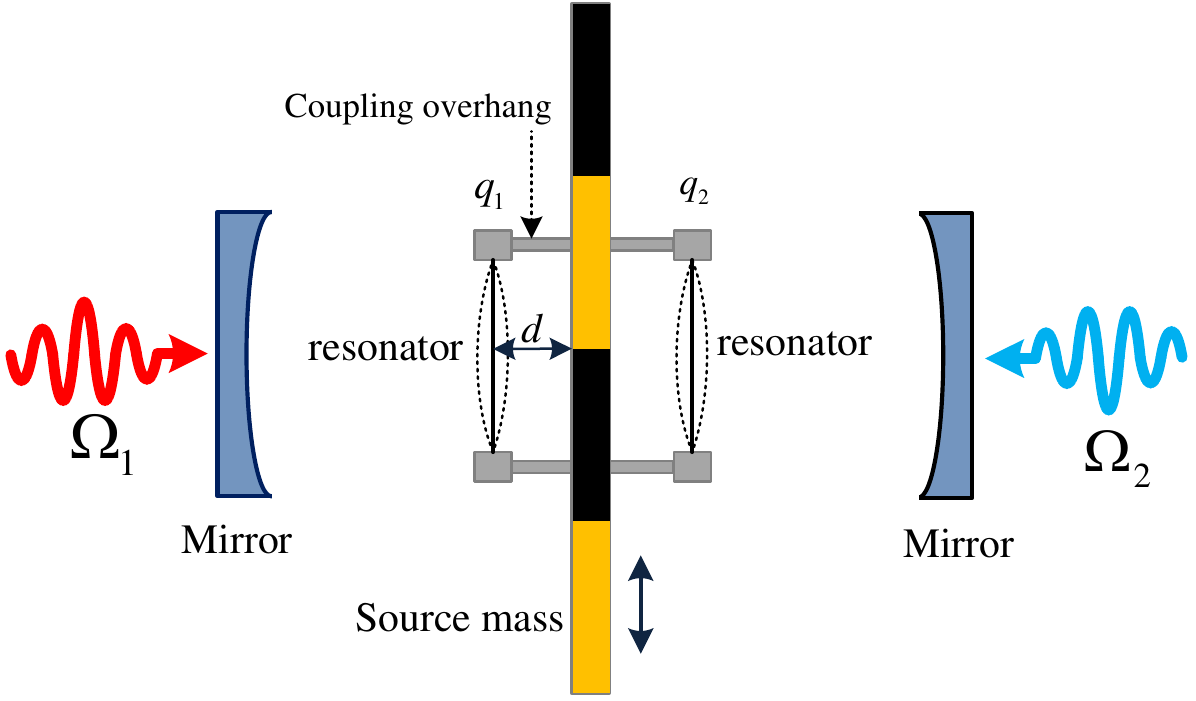}
\caption{Generic setup. A coupled cavity optomechanical system with double
resonators. A source mass with varying density sections is positioned in the
middle. The two cavities are driven by blue- and red-detuned laser,
respectively.}
\end{figure}
Let us consider a coupled cavity optomechanical system with double
resonators. A schematic of our setup is sketched in Fig. 1, where two
mechanical resonators are coupled to two cavities respectively, and
simultaneously coupled to each other. There is a source mass with varying
density sections positioned between the two cavities with the same
separation $d$ from each resonator. The two cavities are driven by blue- and
red-detuned laser fields, respectively. The Hamiltonian of the optical
controlled mechanical gain/loss optomechanical systems can be expressed as ($%
\hbar =1$).%
\begin{equation}
\begin{split}
H& =\underset{j=1,2}{\sum }[\Delta _{j}c_{j}^{+}c_{j}+\omega
_{j}(q_{j}^{2}+p_{j}^{2})-g_{0}c_{j}^{+}c_{j}q_{j}] \\
& -Jq_{1}q_{2}+\underset{j=1,2}{\sum }(\Omega _{j}c_{j}^{+}+H.c.),
\end{split}%
\end{equation}%
where $\omega _{j}$ is the resonance frequency of jth resonator, $\Delta
_{j} $ is the frequency detuning between the jth cavity mode and the jth
external driving field. The first term denotes two cavity modes, two
mechanical modes and the optomechanical coupling $g_{0}$ between them, the
second term corresponds the coupling $J$ between the mechanical resonators
and the third term represents the two driving lasers with strengths $\Omega
_{i}$.

The dynamics of the cavity fields and resonators can be described by quantum
Langevin equations. We can neglect the fluctuations of the cavity fields and
mechanical resonators in the strong external driving condition and from
Eq.(2) we can get the equations of motion for the mechanical modes without
explicit optical coupling%
\begin{equation}
\frac{d}{dt}p_{1}=-(\omega _{1}+\delta \omega _{1})q_{1}+Jq_{2}-(\Gamma
_{1}+\gamma _{1})p_{1},
\end{equation}%
\begin{equation}
\frac{d}{dt}p_{2}=-(\omega _{2}+\delta \omega _{2})q_{2}+Jq_{1}-(\Gamma
_{2}+\gamma _{2})p_{2},
\end{equation}%
where the optically induced mechanical frequency shift and damping rate
are[27],%
\begin{equation}
\begin{split}
\delta \omega _{j}& =2g_{0}^{2}n_{cav}^{(j)}[\frac{\Delta _{j}-\omega _{j}}{%
(\kappa _{j}/2)^{2}+(\Delta _{j}-\omega _{j})^{2}} \\
& +\frac{\Delta _{j}+\omega _{j}}{(\kappa _{j}/2)^{2}+(\Delta _{j}+\omega
_{j})^{2}}],
\end{split}%
\end{equation}%
\begin{equation}
\begin{split}
\Gamma _{j}& =g_{0}^{2}n_{cav}^{(j)}\kappa _{j}[\frac{-1}{(\kappa
_{j}/2)^{2}+(\Delta _{j}-\omega _{j})^{2}} \\
& +\frac{1}{(\kappa _{j}/2)^{2}+(\Delta _{j}+\omega _{j})^{2}}].
\end{split}%
\end{equation}%
Here $n_{cav}^{(j)}$ represent the photon number circulating inside the jth
cavity, $\kappa _{j}$ is the jth overall cavity intensity decay rate. With
the optical controlling on mechanical properties, we can obtain PT
symmetrical systems through the cavity optomechanical effects.

With the simplification, we use two identical optical cavities and
resonators, that means $\omega _{j}(j=1,2)=\omega _{m}$, $\kappa
_{j}(j=1,2)=\kappa $. We use the same driving laser to drive both optical
cavities simultaneously, $\Delta _{j}(j=1,2)=\Delta $, $%
n_{cav}^{(j)}(j=1,2)=n_{cav}$. Thereby $\delta \omega _{j}(j=1,2)=\delta
\omega _{m}$ ,$\Gamma _{j}(j=1,2)=\Gamma _{m}$. When $\gamma _{eff}=-(\Gamma
_{m}+\gamma _{1})=\Gamma _{m}+\gamma _{2}$, one can get the dynamical
equations for the coupled optomechanical resonators with PT symmetry%
\begin{equation}
\frac{d}{dt}q_{1}=\omega _{eff}p_{1},
\end{equation}%
\begin{equation}
\frac{d}{dt}q_{2}=\omega _{eff}p_{2},
\end{equation}%
\begin{equation}
\frac{d}{dt}p_{1}=-\omega _{eff}q_{1}+Jq_{2}+\gamma _{eff}p_{1},
\end{equation}%
\begin{equation}
\frac{d}{dt}p_{2}=-\omega _{eff}q_{2}+Jq_{1}-\gamma _{eff}p_{2},
\end{equation}%
Here the effective resonator frequency $\omega _{eff}=\delta \omega +\omega
_{m}$. Now, let us derive expressions describing the physical properties of
the boundary between the broken- and unbroken-PT -symmetry regions.
Eqs.(7)--(10) can be rewritten in a compact matrix form as%
\begin{equation}
i\frac{d}{dt}\left\vert \Psi \right\rangle =H_{eff}\left\vert \Psi
\right\rangle ,
\end{equation}%
with $\left\vert \Psi \right\rangle =(q_{1},p_{1},q_{2},p_{2})^{T}$, and the
effective Hamiltonian%
\begin{equation}
H_{eff}=i\left(
\begin{array}{cccc}
0 & \omega _{eff} & 0 & 0 \\
-\omega _{eff} & \gamma _{eff} & J & 0 \\
0 & 0 & 0 & \omega _{eff} \\
J & 0 & -\omega _{eff} & -\gamma _{eff}%
\end{array}%
\right) .
\end{equation}%
The eigenvalues of the effective Hamiltonian $H_{eff}$ are given as%
\begin{equation}
\omega _{\pm }=\sqrt{\omega _{eff}^{2}-\frac{\gamma _{eff}^{2}}{2}\pm \sqrt{(%
\frac{\gamma _{eff}^{2}}{2})^{2}+\omega _{eff}^{2}(J^{2}-\gamma _{eff}^{2})}}%
.
\end{equation}

\begin{figure}[tbp]
\includegraphics[width=8cm]{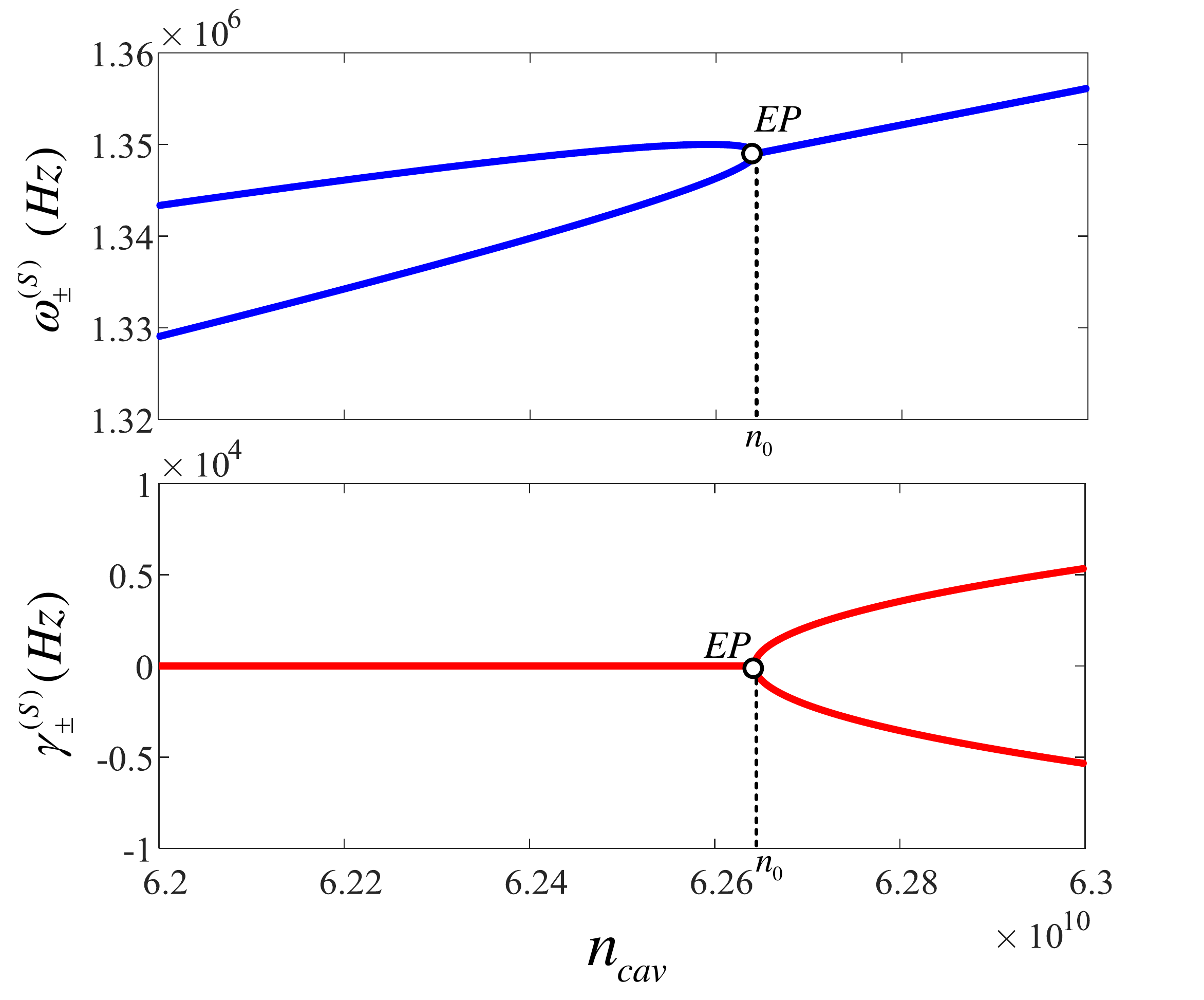}
\caption{Real(left) and imaginary(right) part of the mechanical
eigenfrequency as a function of $n_{cav}$ with $\Delta =\protect\omega _{m}$%
. }
\end{figure}
There are many types of optomenchanical resonators. For illustrating, we
choose one of them, in which the mechanical degree of freedom is a flexible,
partially transparent object (such as a dielectric membrane) placed inside a
Fabry-Perot cavity[28]. Here we use a commercial SiN membrane $1mm$ square
by $50nm$ thick. To bridge the theoretical proposal and experiments, we use
the following practical device parameters $\omega _{m}=100kHz$ $\kappa =10MHz
$, $g_{0}=50Hz$. Tunable coupling of two vibrating mechanical resonators can
be achieved by the piezoelectric effect or the photothermal effect[3] and
the coupling rate $J=\kappa /100=0.1MHz$. We then set the cavity-pump
detuning $\Delta =\omega _{m}$. Fig.2 shows that the PT-broken regime and
the PT symmetric regime are separated by the exception point, where the
eigenfrequencies coalesce. Here $\omega _{\pm }^{(S)}=$Re$(\omega _{\pm })$
and $\gamma _{\pm }^{(S)}=$ Im$(\omega _{\pm })$ correspond to the
frequencies and linewidths of the super modes. As $n_{cav}$ increases, the
frequencies of the pair of super modes approach to each other and coalesce,
while the linewidths starts with zero and then branches, indicating that the
PT symmetry is broken.

\section{MEASUREMENT}

In the following we propose a method for measuring the force gradient. In
such experiments the Casimir effect is an unwanted background that needs to
be suppressed. The \textquotedblleft isoelectronic\textquotedblright\ or
\textquotedblleft Casimir-less\textquotedblright\ technique introduced
firstly in Ref.[29] can be used to achieve this purpose. The mass source
with $20mm$ square by $500nm$ thick is composed of alternating strips of
different materials (e.g., Au and Si) as shown in Fig.1. In the absence of
electromagnetic contributions, a comparison of the forces exerted on a test
mass (resonator) by materials of different densities leads to constraints on
$\alpha $ and $\lambda $ of Eq. (1). When the mass source moves up and down,
the change of forces exerted on the test mass can be regarded as the
perturbation of the optomechanical system. The frequency shift is
proportional to the variation of force gradient at the mass of center of the
resonators.

\begin{equation}
\Delta \omega =-\frac{1}{2m_{t}\omega _{m}}\frac{\partial F}{\partial r}.
\end{equation}%
Thereby for an ordinary sensor, the induced frequency shift is proportional
to the strength of the perturbation $\partial F/\partial r$.

\begin{figure}[tbp]
\includegraphics[width=8.5cm]{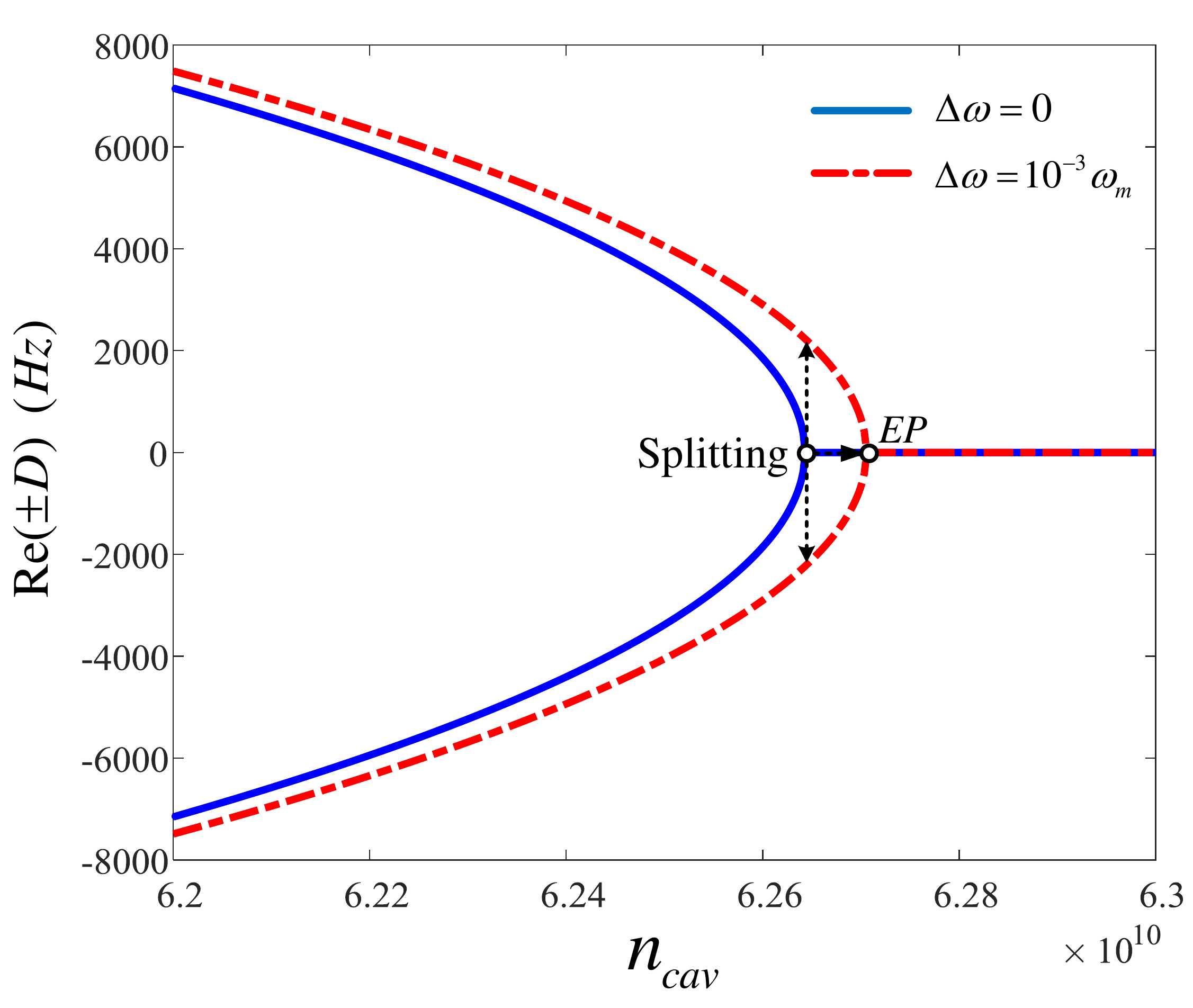}
\caption{Eigenfrequency difference of the effective mechanical system before
perturbation (full blue lines) and after perturbation (dashed red lines) by
variation of force gradient corresponding to a frequency shift of $\Delta
\protect\omega =10^{-3}\protect\omega _{m}$.}
\end{figure}
An important parameter in a PT symmetrical system is the difference between
the eigenfrequencies, namely the eigenfrequency splitting which can be
defined as $D=$Re$(\omega _{+}-\omega _{-})$. We plot $D$ as a function of $%
n_{cav}$ in Fig. 3 by the blue line. At EP, both eigenvalues and
eigenvectors are coalesce. If the system is subjected to a weak perturbation
$\Delta \omega $, the eigenfrequency spectrum will be changed form $D(\omega
_{m})$ to $D(\omega _{m}+\Delta \omega )$. In the same figure, we show the
eigenfrequency of the effective optomechanical system after the perturbation
$\Delta \omega =10^{-3}\omega _{m}$ appears with a dash red line. This
perturbation can be caused by a variation in the force gradient of the
resonators. From the figure we can see that the perturbation pushes the
exceptional point to the right, and therefore the non-Hermitian degeneracy
of the eigenfrequency is released at the exceptional point before the
perturbation, which causes the super mode frequency to split. This splitting
can be used as a signal of the non-Newtonian effect.

\section{SENSITIVITY}

\begin{figure}[tbp]
\includegraphics[width=8cm]{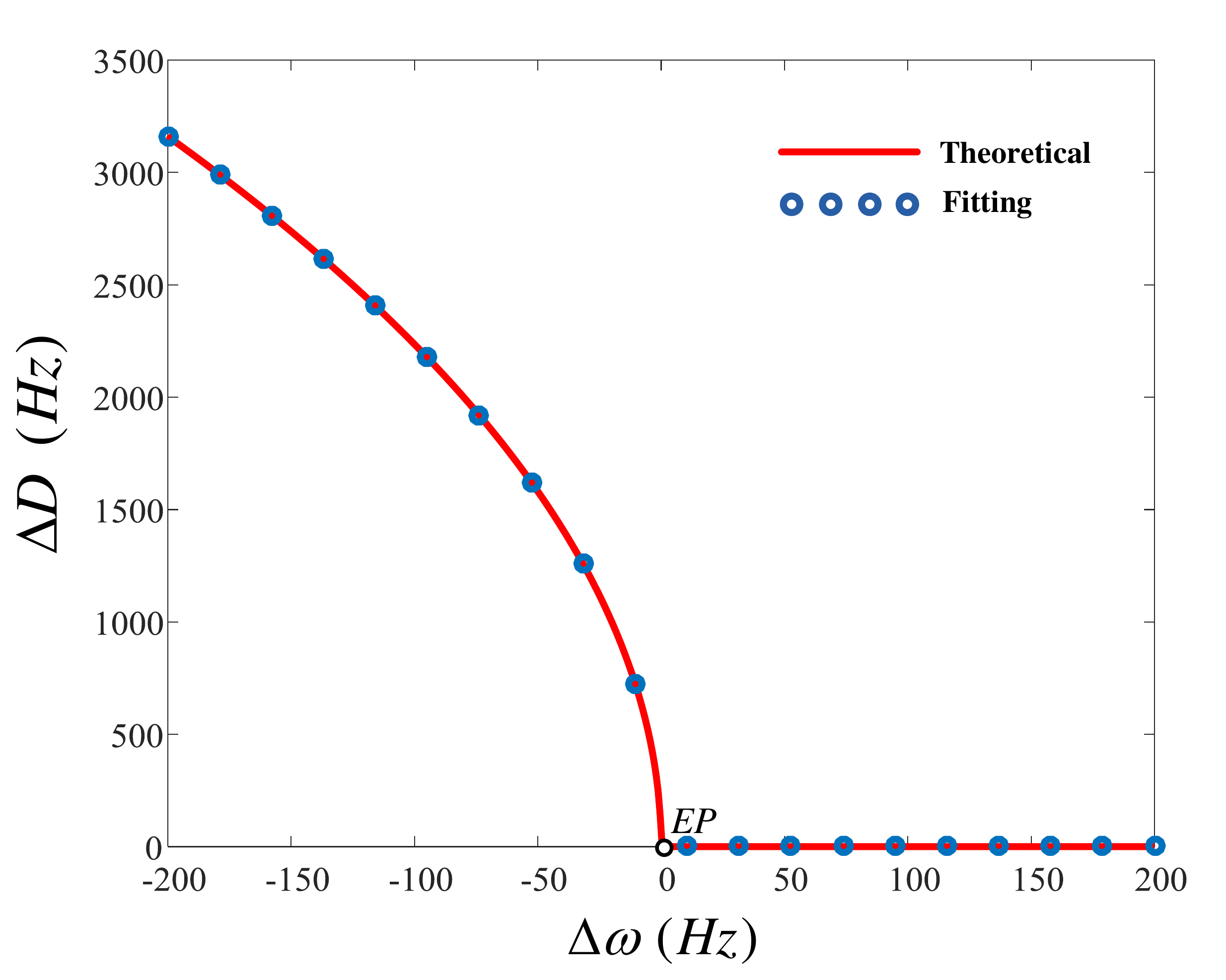}
\caption{Theoretical and fitting curves of super mode frequency splitting as
a function of perturbation. The fitting expression is Eq.(15) with $%
Y=5\times 10^{4}$.}
\end{figure}
When we are trying to evaluate the sensitivity of the sensor at EP, we need
to know the effect of frequency perturbations on the super mode frequency
splitting near the EP ($n_{0}\simeq 6.2643\times 10^{10}$). Therefore, the
sensitivity is defined as $\Delta D(\Delta \omega )=[D(\omega _{m})-D(\omega
_{m}+\Delta \omega )]\mid _{n_{0}}$. Fig.4 shows $\Delta D$ as a function of
the perturbation $\Delta \omega $ at EP. We fit the curve numerically to get
an equivalent expression

\begin{equation}
\Delta D\simeq \sqrt{Y\Delta \omega }.
\end{equation}

Here the constant $Y=5\times 10^{4}$. It is not difficult to see that the
eigenfrequency splitting near the EP point is proportional to the square
root of the frequency perturbation $\Delta \omega $ that is, proportional to
the square root of $\partial F/\partial r$. For a conventional sensor, the
frequency change is linear dependence. Therefore due to the complex square
root topology near the exception point, the eigenfrequency splitting is
greatly enhanced for sufficiently small perturbations compared to sensors
operating under non-exception point sensing schemes.

We can define the sensitivity enhancement factor $\eta $ as the square root
dependence on $\Delta \omega $ instead of a linear dependence, which can be
given by $\eta =\Phi /\Delta \omega =\sqrt{Y/\Delta \omega }$. This equation
shows that the sensitivity enhancement is inversely proportional to the
square-root of the perturbation $\Delta \omega $. So $\eta $ is greatly
increased for weak perturbation, proving the efficiency of the EP sensor in
detecting weak forces.

The resolution of the spectrum is determined by the full width at half
maximum (FWHM) of the spectral lines. The linewidth of the mechanical
resonator in the optomechanical system can be calculated by%
\begin{equation}
\sigma =\frac{\omega _{n}}{Q}.
\end{equation}%
The membrane have been measured $Q=1.2\times 10^{7}$ at $T=300mK$[28], then
we can obtain $\sigma =8.3mHz$, which is the smallest detectable frequency
shift in traditional optomechacial spectrum. At the same time, due to
various noise processes, the spectral lines will broaden[30]. Because the
membrane has a high mechanical quality factor and meets low temperature
conditions, the thermal noise of the system can be controlled at a level
much lower than the line width. On the other hand, under high vacuum
conditions(approximately $10^{-4}Pa$), the momentum exchange noise of the
resonators can also be greatly reduced to a negligible level. The strength
of the perturbation required to cause the same frequency shift in the EP
system is smaller due to the complex square root topology. To compare
ordinary sensors with EP-based sensors, we making $\Delta D=\sigma $ in
Eq.(15). Thus we get $\Delta \omega _{\min }\sim 10^{-9}Hz$ (without EP is $%
8.3mHz$) and then we can obtain the force gradient resolution with Eq.(14), $%
(\partial F/\partial r)_{\min }=2m_{t}\omega _{m}\Delta \omega _{\min }\sim
10^{-14}N/m$. In order to measure the interactions at ultra-short distances,
we bring the membrane close enough to the mass source with a distance of $%
d=100nm$. The mass source we use is a plate with a thickness of $500nm$, so
the distance between the membrane and the center of the plate is $r=375nm$.
In these conditions, the force sensitivity of the EP scheme is at the level
of $10^{-20}N$.

\begin{figure}[tbp]
\includegraphics[width=8cm]{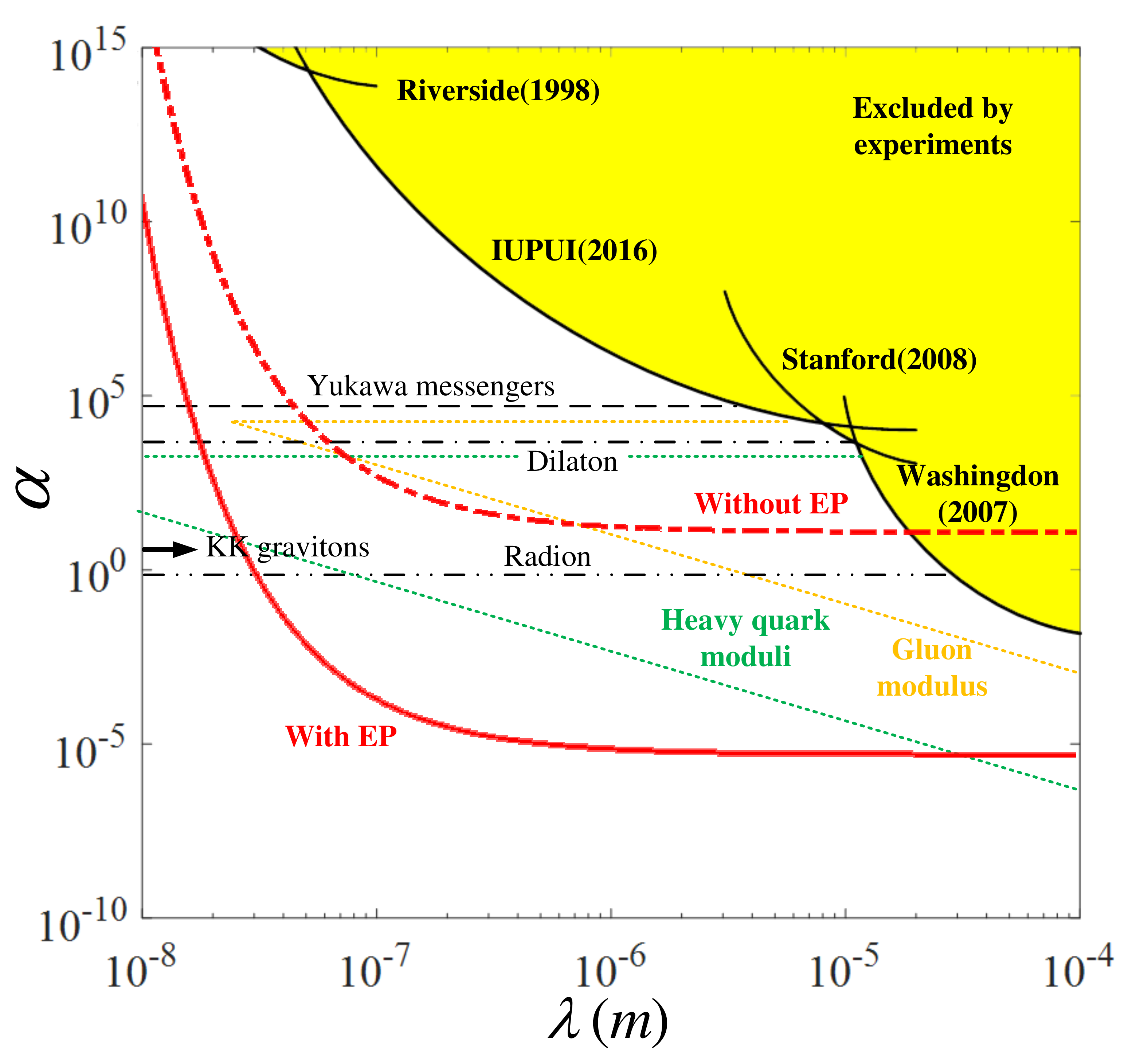}
\caption{Experimental constraints of short-range forces caused by
Yukawa-type interaction potentials [19-22] and theoretical predictions of
this work. The shaded region is excluded with 95\% confidence level. The
full line and dashed denote the results of EP measurement scheme and
conventional optomechanical sensor respectively.}
\end{figure}
In Fig. 5 we describe the experimental constraints and theoretical
prediction on the forces caused by the interaction potential of the Yukawa
form [19-22]. The dashed line is the constraint from the conventional
optomechanical sensor, and the full black line represents the result of the
EP measurement scheme. It can be clearly seen that the EP-based
optomechanical nanogravity gradiometer performs better. We hope that the
proposed system can advance the search for non-Newtonian gravity by an
enhanced sensitivity of $\eta =6.4\times 10^{6}$ compared with traditional
optomechanical method and reach an unprecedented sensitivity for force
sensing.

\section{CONCLUSION}

Based on cavity optomechanics we propose a high-efficiency optomechanical
nano-gravimeter  operating at exceptional points, non-Hermitian
degeneracy. The system consists of two mechanically coupled optomechanical
resonators. The cavity is driven by a blue detuned (red detuned) laser to
induce mechanical gain (loss). The system has an EP at the gain and loss
balance, where any perturbation will cause a frequency split, which is
proportional to the square root of the intensity of the perturbation. Hence,
for a sufficiently small perturbation, the splitting at the exceptional
point is larger compared to traditional optomechanical sensors. This
particular characteristic of exceptional points has been proposed for use in
weak short range force measurement.

\begin{acknowledgments}
This work was supported by the National Natural Science Foundation of China
(Nos.11274230 and 11574206), the Basic Research Program of the Committee of
Science and Technology of Shanghai (No.14JC1491700).
\end{acknowledgments}

\end{document}